\documentclass[showpacs,preprintnumbers,amsmath,amssymb,floatfix,12pt]{revtex4}
\usepackage{graphicx}

\newcommand\figcaption{\def\@captype{figure}\caption}
\newcommand\tabcaption{\def\@captype{table}\caption}
\def\al{\alpha}
\def\vr{\varrho}

\def\pa{\partial}
\def\vf{\varphi}

\def\ga{\gamma}

\def\th{\theta}
\def\sg{\sigma}
\def\N{{\cal N}}
\def\R{{\cal R}}

\def\diag{\mbox {diag}}

\begin{document}
\title{\LARGE Structure of the Star with Ideal Gases}
\date{\today}
\author{Ying-Qiu Gu}
\email{yqgu@fudan.edu.cn}
\affiliation{School of Mathematical
Science, Fudan University, Shanghai 200433, China}

\begin{abstract}
In this paper, we provide a simplified stellar structure model for
ideal gases, in which the particles are only driven by gravity.
According to the model, the structural information of the star can
be roughly solved by the total mass and radius of a star. To get
more accurate results, the model should be modified by introducing
other interaction among particles and rotation of the star.

\end{abstract}

\maketitle

\section{Some phenomena inside a star}
\setcounter{equation}{0}

The spherical symmetric metric for a static star is described by
Schwarzschild metric
\begin{eqnarray}
g_{\mu\nu}= \diag\left( b(r),-a(r), -{r}^{2},-{r}^{2}
  \sin^2  \th\right).\label{1.1}
\end{eqnarray}
For the energy momentum tensor of perfect fluid
\begin{eqnarray}
T_{\mu\nu}=(\rho+P)U_\mu U_\nu -Pg_{\mu\nu}= \diag\left( b \rho ,a
P, {r}^{2} P, {r}^{2}
  \sin^2 \th P\right),\label{1.2}
\end{eqnarray}
where $\rho(r), P(r)$ are proper mass energy density and pressure,
$U_\mu=(\sqrt{b},0,0,0)$, we have the independent equations as
follow
\begin{eqnarray}
\left(\frac r a\right)'&=&1-8\pi G \rho r^2,~~~~\qquad(G_{00}=-8\pi G T_{00}),\label{1.3}\\
\frac { b'} {b}&=&\frac {a-1} r + 8\pi G  P a  r,~~~(G_{11}=-8\pi G T_{11}),\label{1.4}\\
P'&=&-({\rho +P})\frac { b' } {2b},~~~\qquad
(T^{1\nu}_{~;\nu}=0).\label{1.5}
\end{eqnarray}

To determine the stellar structure of an irrotational star, we
solve these equations. However (\ref{1.3})-(\ref{1.4}) is not a
closed system, the solution depends on an equation of state(EOS)
$P=P(\rho)$. For polytropes, we take $P=P_0\rho^\ga$\cite{wein}.
For the compact stars, there are a lot of EOS derived from
particle models\cite{Ntrn1}-\cite{Ntrn6}, which provide the
structural information and parameters such as the maximum mass for
neutron stars. However, in these EOS the boosting effects of the
gravity on particles seem to be overlooked or adopted
unconsciously. As shown in \cite{gu1}, the static equilibrium
equation (\ref{1.5}) is insufficient to describe such dynamical
effects on fluid, and a Gibbs' type law should be included. In
this paper, according to this Gibbs' type law, we derive the
stellar structure model, which shows that the boosting effects of
gravity play an important role to the stellar structure.

Before expanding the model, we make a few simple calculations and
examine the behavior of the metric and particles inside a star to
get some intuition. The first phenomenon is that, the temporal
singularity and spatial singularity occur at different time and
place if the spacetime becomes singular, and the temporal one
seems to occur firstly.

Denoting the mass distribution by
\begin{eqnarray}
M(r)=4 \pi G \int^r_0 \rho r^2 dr,\qquad \R=2M(r).\label{1.6}
\end{eqnarray}
Then by (\ref{1.3}) we have solution
\begin{equation} a=\left \{ \begin{array}{ll}
  \left(1-\frac {\R(r)} r\right)^{-1}, ~~& {\rm if~~} r< R ,\\
 \left(1-\frac {R_s} r\right)^{-1}, ~~& {\rm if~~} r\ge R,
\end{array} \right. \label{1.7}
\end{equation}
where $R$ is the radius of the star, and the Schwarzschild radius
becomes
\begin{eqnarray}
R_s=2M(R)=\R(R)=8 \pi G \int^R_0 \rho r^2 dr.  \label{1.7.1}
\end{eqnarray}
For any normal star with $R>R_s$. From the above solution we learn
$a(r)$ is a continuous function and
\begin{eqnarray}
a\ge 1, ~~~a(0)=\lim_{r\to\infty} a=1,~~~a_{\rm
max}=a(r_m),~(0<r_m\le R).\label{1.8}
\end{eqnarray}
So the spatial singularity $a\to \infty$ does not appear at the
center of the star when the singularity begins to form.
\begin{figure}[t]
\centering
\includegraphics[width=12cm]{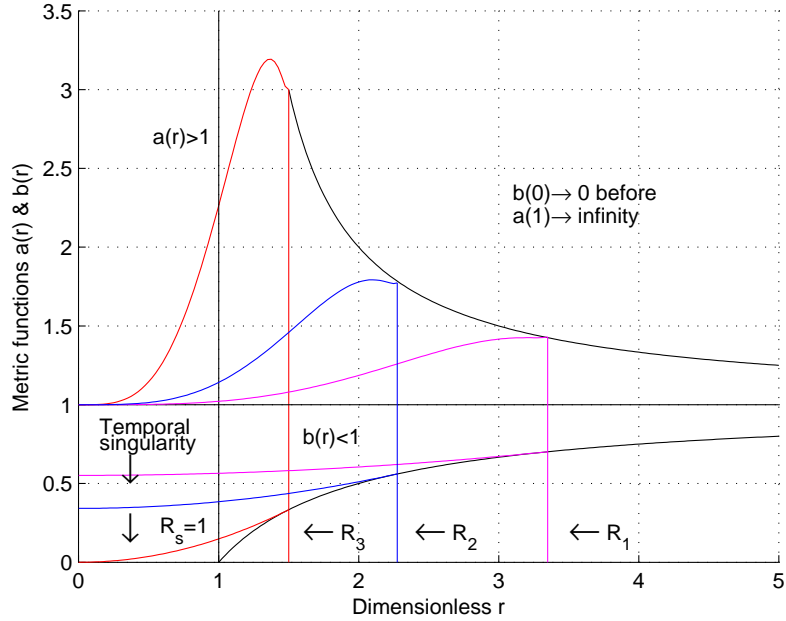}
\caption{The trends of $(a, b)$ as $R\to R_s$, which show the
spatial singularity does not occur at the center, but temporal
singularity may occur at the center before $a\to \infty$}
\label{fig1}
\end{figure}

On the other hand, by (\ref{1.4}) and $P\ge 0$, we find $b'(r)$ is
a continuous function satisfying
\begin{eqnarray}
b'(0)=0, \quad b'(r)>0, ~(\forall r>0), \quad b=1-\frac{R_s}
r,~~(r\ge R).\label{1.9}
\end{eqnarray}
(1.9) shows $b(r)$ is a monotone increasing function of $r$ with
smoothness at least $C^1([0,\infty))$. Consequently, the temporal
singularity $b\to 0$ should take place at the center.

The trends of $a(r),~b(r)$ are shown in FIG.\ref{fig1}, where we
take the Schwarzschild radius $R_s=1$ as length unit. From
FIG.\ref{fig1} and some simplified calculations, we find $b(0) \to
0$ seems to appear before the space becomes singular $a\to
\infty$.

The second phenomenon is that, the particles near the center of
the star are unbalanced, and violent explosion takes place inside
the star before the temporal singularity occurs. When $b(0)\to 0$,
by (\ref{1.9}) we have
\begin{eqnarray}
b \to b_0 r^\al,~~(\al>1),\qquad {\rm if}~~ r \ll R.\label{1.10}
\end{eqnarray}
Substituting it into (\ref{1.5}), we find
\begin{eqnarray}
-P'\to  (\rho +P)\frac \al{2r}\to +\infty,\quad (r\to 0).
\label{1.11}
\end{eqnarray}
According to fluid mechanics, $-\pa_r P$ corresponds to the radial
boosting force, so (\ref{1.11}) means violent explosion.

More clearly, we examine the motion of a particle inside the star.
Solving the geodesic in the orthogonal subspace $(t, r, \th)$, we
get\cite{gu2,gu3}
\begin{eqnarray}
\dot t=\frac {1}{C_1 b},~~\dot \th=\frac{C_2}{r^2},~~\dot\vf=0,
~~\dot r^2=\frac 1 r \left( \frac {1}{C^2_1
b}-\frac{C^2_2}{r^2}-1\right), \label{1.12}
\end{eqnarray}
where $C_1, C_2$ are constants. The normal velocity of the
particle is given by
\begin{eqnarray}
v^2_r=\frac{a dr^2}{b dt^2}=1- {C^2_1}b\left(1+\frac
{C^2_2}{r^2}\right),~~ v^2_\th=\frac{r^2d\th^2}{b
dt^2}=\frac{C_1^2 C^2_2 b}{r^2},~~v^2_\vf=0.  \label{1.13}
\end{eqnarray}
The sum of the speeds provides an equality similar to the energy
conservation law
\begin{eqnarray}
v^2 =1- {C_1^2} b(r),\quad{\rm with}~~~ v^2\equiv
v^2_r+v^2_\th+v^2_\vf. \label{1.14}
\end{eqnarray}
(\ref{1.14}) holds for all particles with $v_\vf\ne 0$ due to the
symmetry of the spacetime.

From (\ref{1.14}) we learn $v\to 1$ when $b\to 0$, this means all
particles escape at light velocity when the temporal singularity
occurs. So instead of a final collapse, the fate of a star with
heavy mass may be explosion and disintegration. The gravity of a
star drives the inside particles to move rapidly and leads to high
temperature. How the particles to react to the collapse of a star
needs further research with dynamical models. A heuristic
computation for axisymmetrical collapse is presented in
\cite{collapse}, which reveals that the fate of a collapsing star
sensitively depends on the parameters in the EOS.

\section{The equations for stellar structure}
\setcounter{equation}{0}

In this paper, we simply take the star as a ball of ideal gases,
which satisfies the following assumptions:

(A1) All particles are classical ones only driven by the gravity,
namely, they are characterized by 4-vector momentum $p_k^\mu$ and
move along geodesic.

(A2) The collisions among particles are elastic, and then they can
be ignored in statistical sense\cite{gu2,gu3}.

(A3) The nuclear reaction and radiation are stable and slowly
varying process in a normal star, in contrast with the mass
energy, the energy related to this process is small noise, so we
treat all photons as particles and omit the process of its
generation and radiation.

These are some usual assumptions suitable for fluid stars.
However, together with (\ref{1.3})-(\ref{1.4}), they are enough to
give us a simplified self consistent stellar structure theory. In
\cite{gu1}, we derived the EOS for such system as follows
\begin{eqnarray}
\N&=& {\N_0}
 \left[{J(J+2\sg)}\right]^{\frac 3 2}, \qquad  J\equiv \frac {kT}{\bar m c^2},\label{2.1}\\
\rho &=& \N \left( \bar m c^2+\frac {3}{2}kT\right)=\vr
 \left[{J(J+2\sg)}\right]^{\frac 3 2}\left( 1+\frac {3}{2}J\right) c^2  , \label{2.2}\\
 P&=&\N kT
\frac{2\sg\bar m c^2+kT}{2(\sg\bar m c^2+kT)}=\vr
\left[{J(J+2\sg)}\right]^{\frac 5 2} \frac{c^2}{2(\sg+J)},
\label{2.3}
\end{eqnarray}
where $\N$ is the number density of particles, $\N_0=\N_0(\bar m,
\sg)$ is related to property of the particles but independent of
$J$, $c=2.99\times 10^8$m/s the light velocity, $\sg\dot = \frac 2
5$ a factor reflecting the energy distribution function, $\bar m$
the mean static mass of all particles, $J$ dimensionless
temperature, which is used as independent variable, $(\rho, P)$
are usual energy density and pressure, $\vr$ a mass density
defined by
\begin{eqnarray}
\vr\equiv \N_0 \bar m.\label{2.4} \end{eqnarray} By (\ref{2.2})
and (\ref{2.3}), we get the polytropic index $\ga$ is not a
constant for large range of $T$ satisfying $1<\ga<\frac 5 3 $, and
the velocity of sound
\begin{eqnarray}
C_{\rm sound}\equiv c \sqrt{\frac {dP}{d\rho}}=\frac {\sqrt 3}
3\left(\frac{c^2J(2\sg+J)(5\sg^2+8\sg
J+4J^2)}{(\sg+J)^2[2\sg+(2+5\sg) J+4J^2]}\right)^{\frac 1 2}<\frac
{\sqrt 3} 3c,\label{2.4.1}
\end{eqnarray}
which shows that the EOS is regular and the causality condition
holds.

Equation (\ref{1.5}) can be rewritten as
\begin{eqnarray}
\frac {d b}{  dJ}=-\frac {2b} {\rho  +P}\frac { d P }
{dJ}.\label{2.5}
\end{eqnarray}
Substituting (\ref{2.2}) and (\ref{2.3}) into (\ref{2.5}), we
solve
\begin{eqnarray}
b = \frac
{4(R-R_s)(\sg+J)^2}{R\left[2\sg+(2+5\sg)J+4J^2\right]^2},
\label{2.6}\end{eqnarray} where $R$ is the radius of the star, and
$R_s=2M_{\rm tot}$ the Schwarzschild radius. Substituting
(\ref{1.7}) and (\ref{2.2})-(\ref{2.6}) into (\ref{1.3}) and
(\ref{1.4}), instead of the Tolman-Oppenheimer-Volkoff equation
and Lane-Emden equation\cite{wein}, we get the following
dimensionless equations for stellar structure,
\begin{eqnarray}
\R'(r) &=& \left(\frac r \chi \right)^2 (2+3J)[J(J+2\sg)]^{\frac 3 2}, \label{2.7}\\
J'(r) &=&\frac{-\xi(J)}{2(r-\R)}\left(\left(\frac r \chi \right)^2
[J(J+2\sg)]^{\frac 5 2}+\frac {\R} r (\sg+J) \right),  \label{2.8}
\end{eqnarray}
where $\chi$ is a constant length scale,
\begin{eqnarray}
\chi&=&c(4\pi G\vr)^{-\frac 1 2}=c(4\pi G\N_0 \bar m)^{-\frac 1 2}, \label{2.9}\\
\xi &\equiv & \frac {2\sg+(2+5\sg)J+4J^2}{5\sg^2+8\sg
J+4J^2},~(\xi \approx 1). \label{2.10}
\end{eqnarray}

\begin{figure}[t]
\centering
  \begin{minipage}[b]{0.4\textwidth}
    \centering
    \includegraphics[width=75mm]{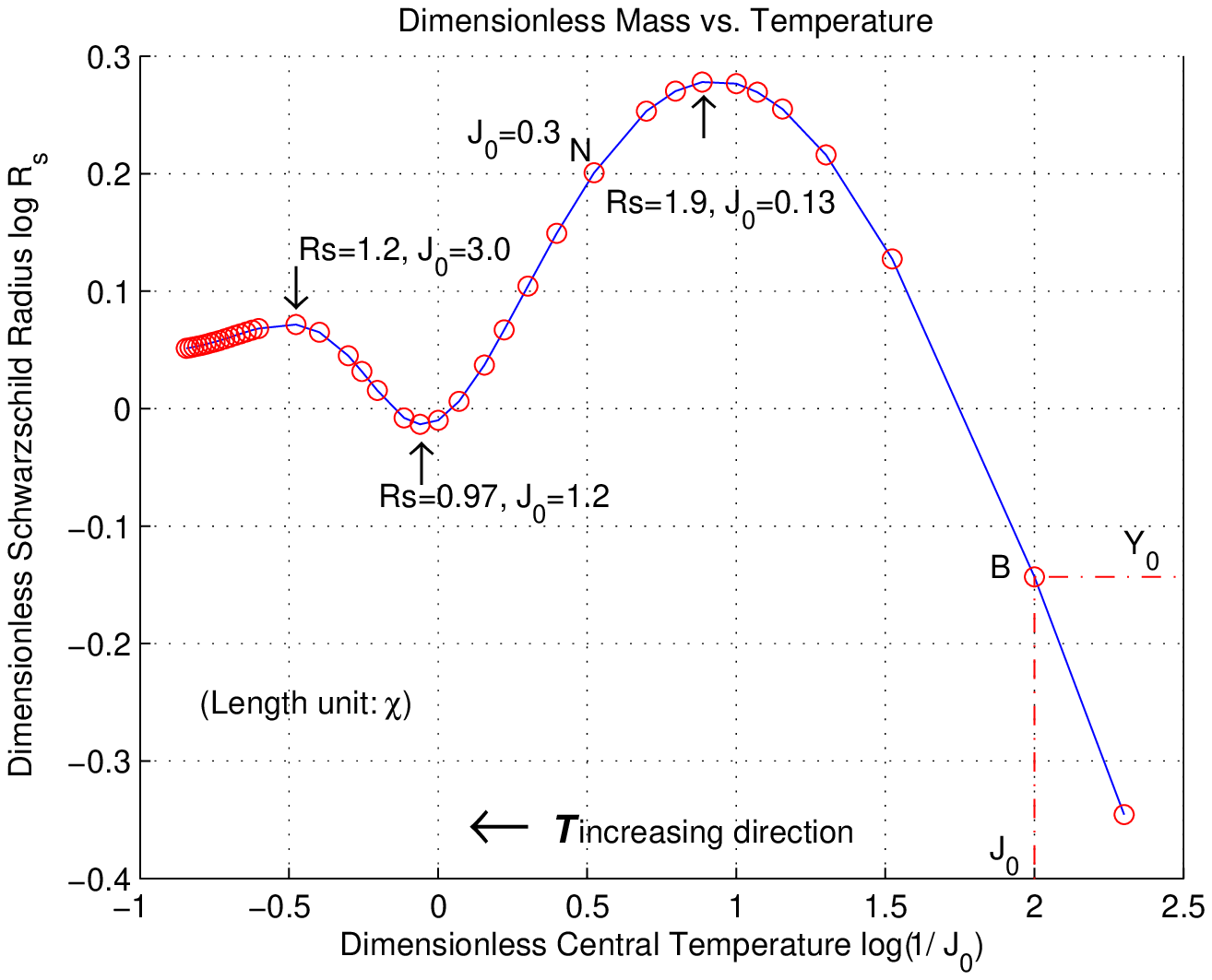}
    \caption{Relation between mass and central
    temperature similar to H-R diagram,
where it is concentrated by the scale $\chi$ and $J$.}\label{fig2}
  \end{minipage}%
  \hspace{0.06\textwidth}%
  \begin{minipage}[b]{0.4\textwidth}
    \centering
    \includegraphics[width=75mm]{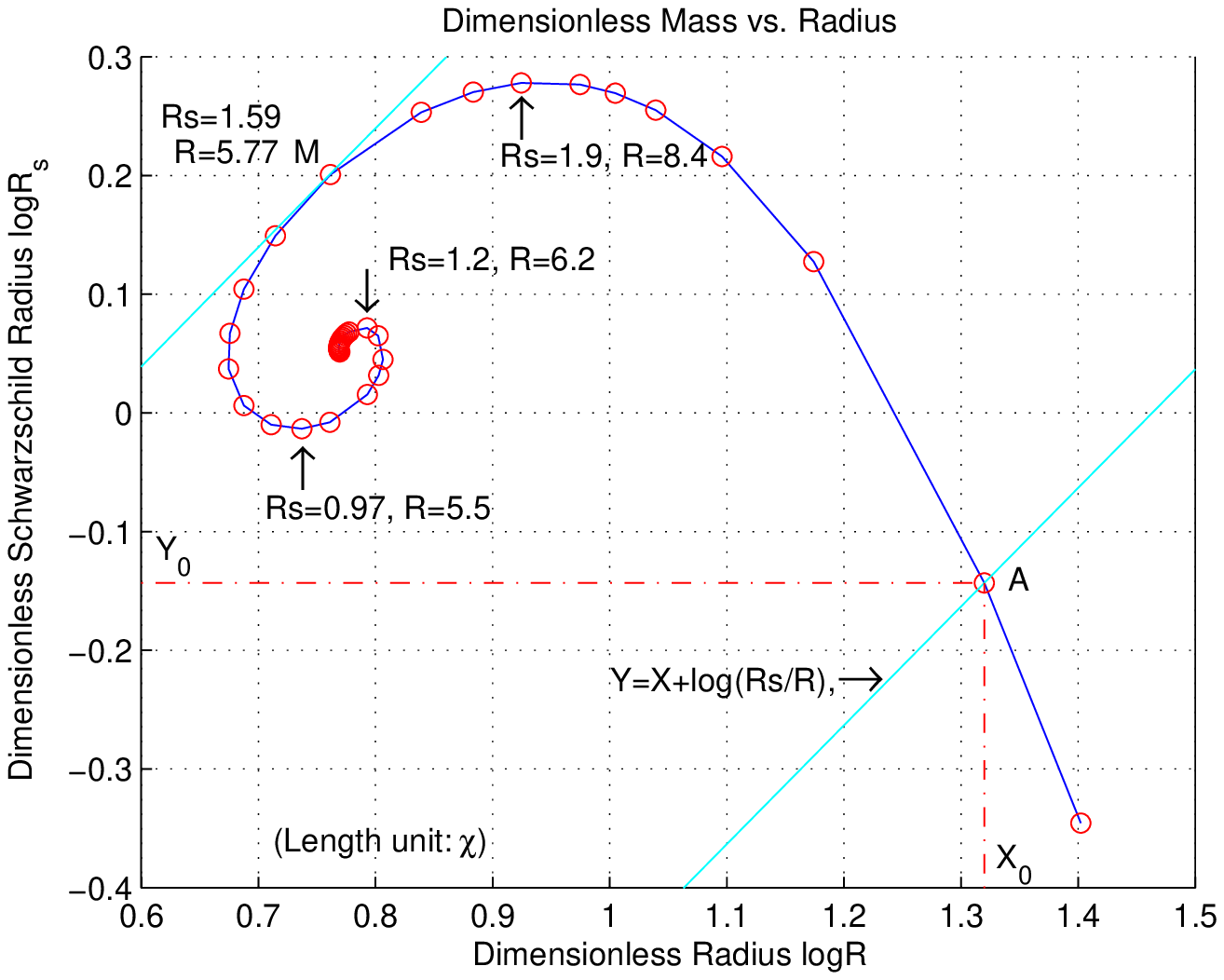}
    \caption{Relation between mass and Radius. All structural
information of a star are determined by a given radii pair $(R_s,
    R)$.}\label{fig3}
  \end{minipage}\\[15pt]
\end{figure}

\begin{figure}[t]
  \centering
  \begin{minipage}[b]{0.4\textwidth}
    \centering
    \includegraphics[width=75mm]{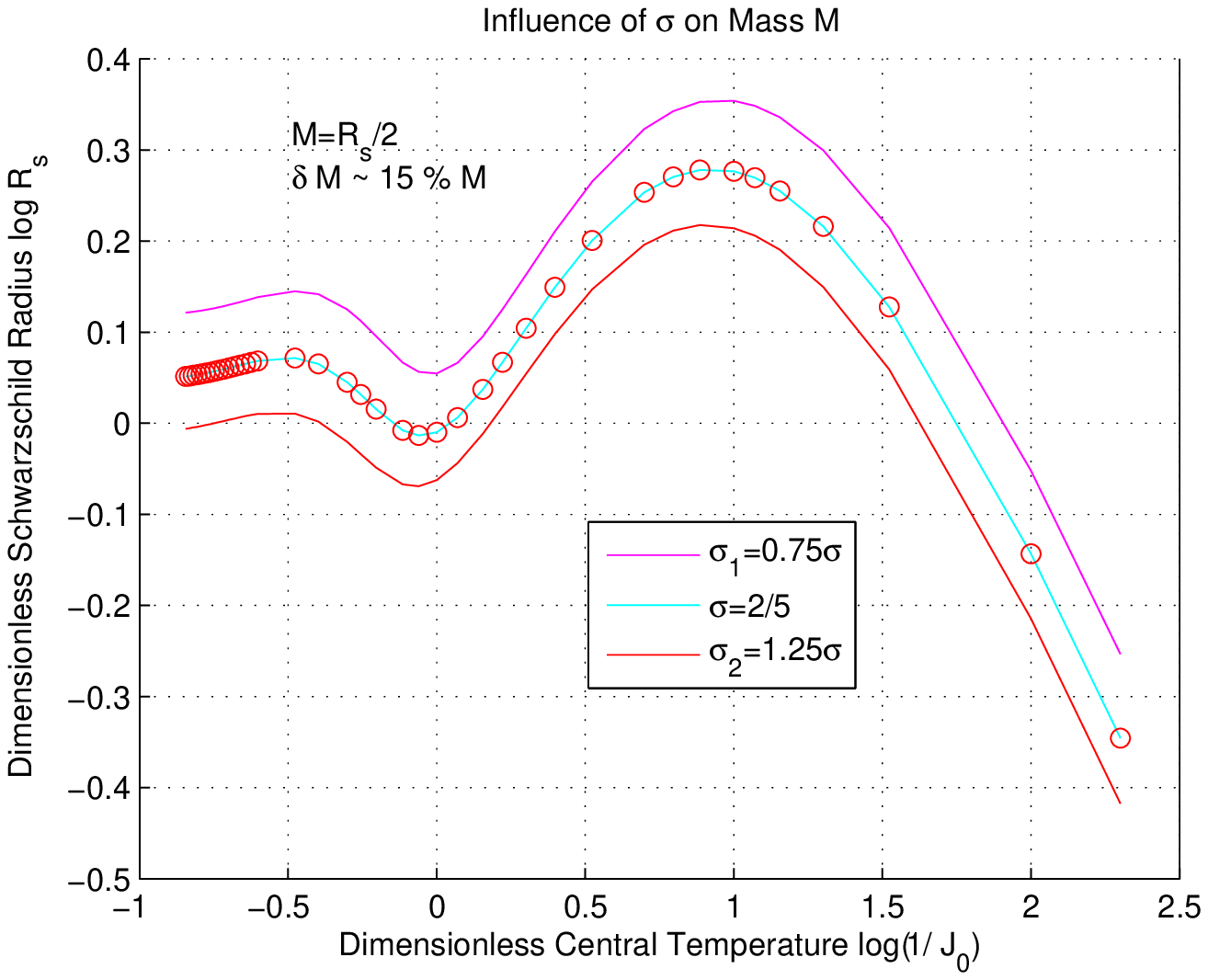}
    \caption{The influence of energy distribution on solutions.
    The results are not sensitive to $\sg$. }\label{fig4}
  \end{minipage}%
  \hspace{0.06\linewidth}%
  \begin{minipage}[b]{0.4\textwidth}
    \centering
    \includegraphics[width=75mm]{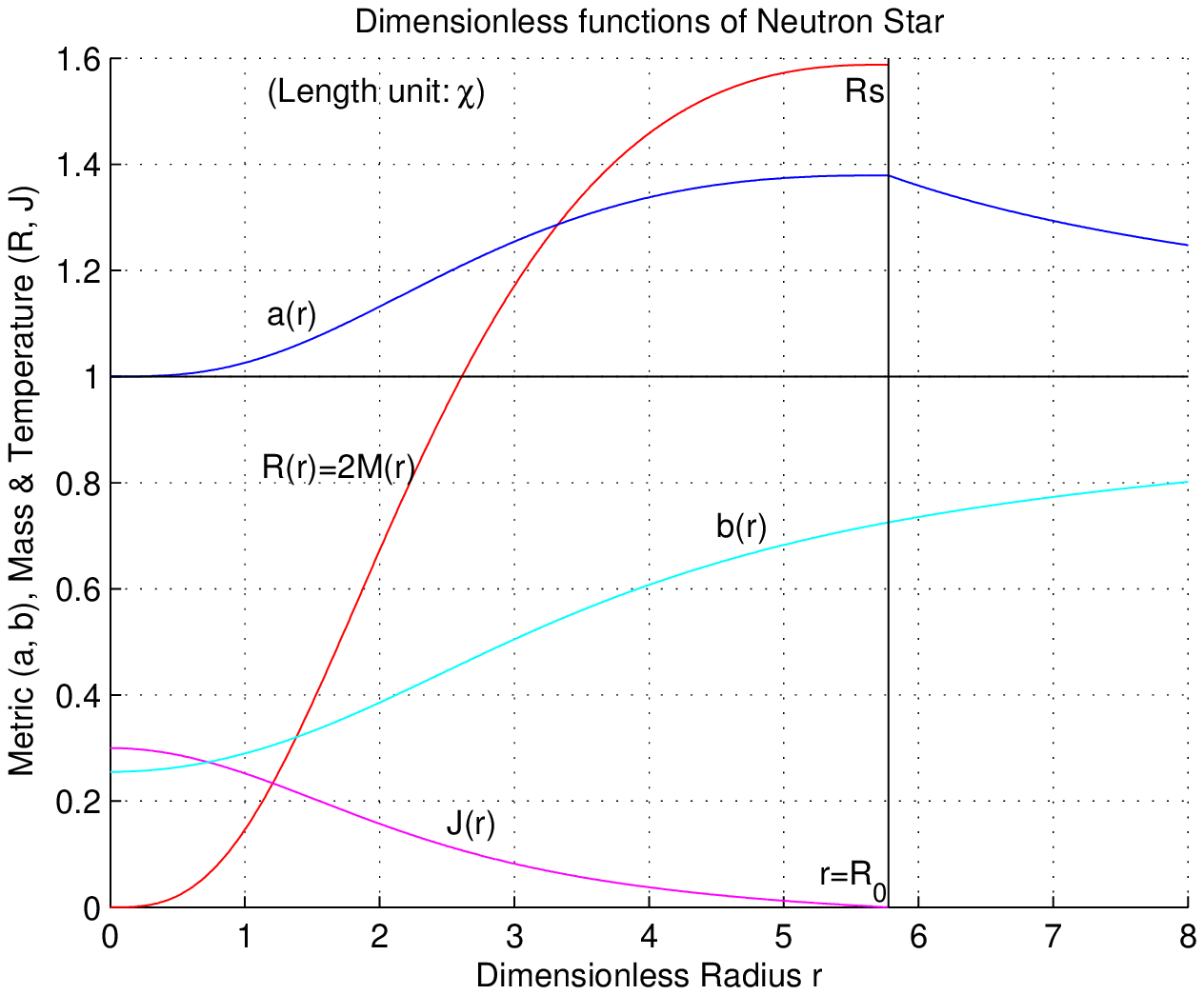}
    \caption{Structural functions for the compactest stars.
    The trends are typical for all stars. }\label{fig5}
  \end{minipage}
\end{figure}

The exact solution to (\ref{2.7}) and (\ref{2.8}) seems not easy
to be obtained. However, they are dimensionless equations
convenient for numeric resolution. If we take $\chi=1$ as the unit
of length, the solution can be uniquely determined by the
following boundary conditions
\begin{eqnarray}
\R(0)=0,~~~J(0)=J_0>0,~~~J(r)=0,~(\forall r\ge R). \label{2.11}
\end{eqnarray}
Adjusting $J_0$, we get different solution. Some solutions are
displayed in FIG.\ref{fig2}-FIG.\ref{fig5}.
Usually we can easily measure the radius $R$ and the total mass or
equivalent Schwarzschild radius $R_s$ of a star. Then the other
structural parameters can be determined by this radii pair $(R_s,
R)$. In what follows, we show how to use FIG.\ref{fig2} and
FIG.\ref{fig3} to solve practical problems.

For the sun, we have the radii pair as
\begin{eqnarray}
R_s=2.96\times 10^3 ~{\rm m},\qquad R_\odot=6.96\times 10^8 ~{\rm
m}.\label{2.12}
\end{eqnarray}
By ${R_s} / R_\odot=4.25\times 10^{-6}$, according to the relation
shown in FIG.\ref{fig3}, we can solve the intersection $A$ and get
the dimensionless radii pair
\begin{eqnarray}
X_0=\log(R/\chi)=2.320,~~~Y_0=\log(R_s/\chi)=-3.052.\label{2.12.1}
\end{eqnarray}
Consequently, we have
\begin{eqnarray}
\chi=10^{-2.32}R=3.335\times 10^6 ~{\rm m}. \label{2.13}
\end{eqnarray}
By $Y_0$ we get intersection $B$ in FIG.\ref{fig2}, and then get
the central dimensionless temperature $J_0=1.145\times 10^{-6}$
for the sun. Taking it as initial value we can solve (\ref{2.7})
and (\ref{2.8}), and then get detailed structural information for
the sun.

By (\ref{2.9}) and (\ref{2.13}), we get
\begin{eqnarray}
\vr=\frac {c^2}{4\pi G\chi^2}= 9.640\times 10^{12} ~({\rm
kg/m}^{3}). \label{2.14}
\end{eqnarray}
Then by (\ref{2.2}) and (\ref{2.3}) we solve the mass density and
pressure at the center
\begin{eqnarray}
\rho(0)&=&\vr [J_0(J_0+2\sg)]^{\frac 3 2}\left(1+\frac 3 2
J_0\right)=8.45\times 10^3
~({\rm kg/m^3}), \label{2.15}\\
P(0)&=&\frac 1 2 \vr [J_0(J_0+2\sg)]^{\frac 5
2}(\sg+J_0)^{-1}c^2=8.70\times 10^{8}~ ({\rm MPa}). \label{2.16}
\end{eqnarray}
The temperature depends on the mean mass $\bar m$. By (\ref{2.1}),
we have
\begin{eqnarray}
T_c={\bar m c^2 J_0}/k=n_p (m_pc^2 J_0/k)=1.247\times 10^7
n_p~{\rm (K)}, \label{2.17}\end{eqnarray} where $m_p=1.673 \times
10^{-27}$kg is the static mass of proton, $n_p$ is the equivalent
proton number for the particles.

If the ionization in the sun is about $H^+ + N^+ + 2e^-$, then we
have
\begin{eqnarray}
n_p&=&(70\%\times 1 +30 \% \times 14)/4=1.23,\\\bar m &=& n_p
m_p=2.06\times 10^{-27}~({\rm kg}),\\
T_c&=&1.247\times 10^7 n_p=1.53\times 10^7 ~{\rm (K)},\\
\N_0 &=&  {c^2}/(4\pi G\bar m\chi^2)=4.68\times 10^{39}~({\rm
m}^{-3}). \label{2.19}
\end{eqnarray}

The compactest stars(with the maximum $R_s/R$) correspond to the
points $M, N$ in FIG.\ref{fig3} and FIG.\ref{fig2}. The radii pair
is $R_s:R=1.59:5.77$ and the central temperature $J_0=0.30$. For a
compact star with the solar mass $M_\odot$, we get the length unit
and radius as
\begin{eqnarray}
\chi=R_s/1.59=1.86~{\rm km},~~~R=5.77\chi=10.7~{\rm km}.
\label{2.20}
\end{eqnarray}
Along the above procedure, we solve
\begin{eqnarray}
\rho(0)&=&8.50\times 10^{18}~{\rm kg/m}^{3},\\
P(0)&=& 1.24\times 10^{29}~{\rm MPa},\\
T(0)&=& 3.27\times 10^{12}n_p~{\rm K}. \label{2.21}
\end{eqnarray}
These are typical data for a neutron star\cite{wein}-\cite{Ntrn6}.
The metric functions $(a, b)$ and the mass, temperature
distributions $(\R, J)$ for this star are displayed in
FIG.\ref{fig5}.

\section{Discussion and conclusion}
\setcounter{equation}{0}

The above numerical results show that, although the orbits of the
particles are not simple ellipses due to collisions, the influence
of the gravity still exists and leads to high temperature inside
the star. The main part of the EOS may be related to gravity.

In contrast with (\ref{2.15}) and (\ref{2.16}), we find the
central density and pressure in the sun are about one order of
magnitude less than the current data
\begin{eqnarray}
\rho(0)=1.6\times 10^5~{\rm kg/m}^3,\qquad P(0)=2.5\times
10^{10}~{\rm MPa}.
\end{eqnarray}
This difference is an unsettled serious problem, which seems to be
caused by the dynamical effect of gravity.

FIG.\ref{fig4} shows that the solutions are not very sensitive to
the concrete energy distribution or $\sg$, so one needs not to
solve specific problems via calculating complex distribution
functions\cite{gu1}.

For photons, by the Stefan-Boltzmann's law
\begin{eqnarray}
\rho=\frac {8\pi^5(kT)^4}{15(hc)^3},
\end{eqnarray}
we get
\begin{eqnarray}
\N_0\to \frac {16\pi^5}{45}\left(\frac {\bar m c} h\right)^3=\frac
{2\pi^2}{45}\left(\frac {\bar m c} \hbar\right)^3,\qquad (\bar
m\to 0).
\end{eqnarray}
How to determine the concrete function $\N_0(\bar m, \sg)$ is an
interesting problem.

The dimensionless equations (\ref{2.7}) and (\ref{2.8}) simplify
the relations between parameters. This function is similar to that
of the similarity theory \cite{similar}. However in these
equations the information of the interaction among particles is
ignored, so it can not provide the critical data such as the
maximum density, the largest mass. To get such data the potentials
and interactive fields should be introduced to the energy momentum
tensor\cite{gu1}.

\newpage
\section*{Acknowledgments}

The author is grateful to his supervisor Prof. Ta-Tsien Li for his
encouragement and guidance.  Thanks to Prof. Jia-Xing Hong for
kind help.

\end{document}